# String Dynamics, Spontaneous Breaking of Supersymmetry and Dual Scalar Field Theory

Lu-Xin Liu

*Department of Physics*
*Purdue University*
*West Lafayette, IN 47907, USA*
liul@physics.purdue.edu

**Abstract**. The dynamics of a vortex string, which describes the Nambu-Goldtone modes of the spontaneous breakdown of the target space D=4, N=1 supersymmetry and internal $U(1)_R$ symmetry to the world sheet ISO(1,1) symmetry, is constructed by using the approach of nonlinear realization. The resulting action describing the low energy oscillations of the string into the covolume (super)space is found to have an invariant synthesis form of the Akulov-Volkov and Nambu-Goto actions. Its dual scalar field action is obtained by means of introducing two vectorial Lagrangian multipliers into the action of the string.

# I. Introduction

Topological defects can form as a result of some types of symmetry breaking if the system has non-trivial degenerate vacua. Examples include the formation of domain walls if the vacuum manifold $M$ has disconnected components [1]; the formation of string topological defects, such as the Nielsen-Olesen vortex lines, if $M$ is not simply connected [2]. In elementary particle physics, the known symmetries of these particles are actually originated from a larger symmetry group $G$ after a series of spontaneous symmetry breakings. In the cosmological context, it implies that the early universe has gone through a number of phase transitions, with one or more types of topological defects, such as the cosmic string, possibly having formed and being left behind [3].

Vortex string topological defects and their theory have attracted much attention, in particular on the structure of the vortex moduli space for non-abelian as well as abelian theories [2,4,5,6,7,8]. In brane world scenarios, particular attention has been focused on the vortex string itself, which is deemed to be especially useful when considering localization of gauge fields by using warped compactifications [9]. In addition, in Ref.[5], it is pointed out that the low energy dynamics of the vortex string in certain field configurations coincides with a two dimensional field theory. It has been found that the vortex string theory has a rich content and is interesting in its own right from both theoretical and aesthetic points of view. On these points the dynamics of the vortex string itself and its correlations with lower dimensional field theories merit careful investigation and analysis.

It has been shown that a supersymmetric string topological defect can form in N=1 supersymmetric abelian Higgs models [7,10,11]. However, recently, a non-BPS solitonic string was found in an N=1 supersymmetric gauge theory [12]. There, the vacuum spontaneously breaks the total N=1 supersymmetry. Motivated by this, in the present paper we consider such a vortex topological defect (or p=1 brane), embedded in D=4 flat target space with the underlying theory described by the global N=1 supersymmetry along with an internal $U(1)_R$ symmetry. Corresponding to the breaking of the (super)spacetime, the dynamics of the vortex string is then described by its long wave oscillation modes in the target (super)space.

Besides, it is known that the target space N=1 supersymmetry can be partially broken down to (0,2) supersymmetry on the world sheet of the string[13], and the resulting string dynamics is described by the Green-Schwarz superstring theory [14]. On the other hand, if one further introduces spinor degrees of freedom on the world sheet of the string, the supersymmetry can be realized as (1,1) or (1,0) instead, which leads to Neveu-Schwarz and heterotic string theories respectively [15]. Some discussions about extended versions of two dimensional supersymmetry can be found in Ref.[16]. In addition, other patterns of partially spontaneous breakdown of two dimensional supersymmetries have been explored in [17]. However, since there is no observation for the superpartners for all the particles in the standard model of particle physics, the supersymmetry must be a broken one. In this regard, it is prominent to notice the non-BPS string admitted by the N=1 supersymmetry theory in [12]. Therefore, in this context we consider the Nambu-Goldstone modes of such an oscillating vortex string, which correspond to the spontaneous breakdown of the target space full N=1 supersymmetry to D=2 ISO(1,1) symmetry on its world sheet. The $R$ symmetry could also become a broken one, and this has been realized in both the dynamical and the spontaneous supersymmetry breaking theories [18-20].

As for the spontaneous symmetry breaking, the approach of nonlinear realization has been demonstrated a natural, economical and elegant framework for treating it. Actually, this method has been applied to a wide range of physical problems most notably in the form of nonlinear sigma models, supersymmetry and brane theories [13, 21, 22]. There, the Lagrangian is invariant with respect to the transformations of some continuous group $G$, but the ground state is not an invariant of $G$ but only of some subgroup $H$. In this context, the resulting phenomenological Lagrangian becomes an effective theory at energies far below the scale of spontaneous symmetry breaking. Consequently, the effective action can be expressed in terms of the dynamics of the Nambu-Goldstone fields.

In this paper, setting aside the Green-Schwarz formalism [13, 14], we put special emphasis on the string dynamics from the point of view of an effective two dimensional field theory and construct its dual form. The purpose of the research work has two respects. The non-BPS vortex string totally breaks the target space supersymmetry, and the symmetry breaking becomes manifest in terms of the inhomogeneous transformations of the Goldstone (Goldstino) fields through the approach of nonlinear realization. As a result, the Maurer-Cartan one-forms naturally give us the dynamics describing its oscillations into the codimensional (super) space. On the other hand, the dual form of the string action is constructed. We hope that it could shed some light on the correlations between the vortex strings resulting from embedded higher dimensional target spacetime and two dimensional field theories.

The organization of this paper is as follows. In section II, the subgroup $SO(1,1) \times SO(2)$ is taken as the stability group. The string topological defect therefore totally breaks the target spacetime supersymmetry and $R$ symmetry. The centrally extended N=(2,2) supersymmetric algebra is constructed through dimension reduction, whileas the transformations of the Nambu-Goldstone (Goldstino) fields are derived explicitly through the Coset structure by using the approach of nonlinear realization. In section III, the low energy effective action of the oscillating vortex string is obtained by means of Maurer-Cartan one-forms. Its dynamics, according to the interplay of the nonlinear realization and brane dynamics [13, 22], is described by the associated Nambu-Goldstone (Goldstino) modes corresponding to the collective degrees of freedom of the Coset manifold. In addition to the transverse long wave oscillating modes associated with the broken generators in the (super) space directions, the string dynamics also provides accommodation for the NG mode related to the broken $R$ symmetry. The action is found to have a synthesis form of the Akulov-Volkov and Nambu-Goto actions:

$$\Gamma_0 = -T\int d^2 x \sqrt{|\det g|} = -T\int d^3 x \det \hat{e} \sqrt{\det(\delta_{ij} - \nabla_A \phi_i \nabla^A \phi_j)} \tag{1}$$

in which $\det \hat{e} = \det \hat{e}_a^{\ B}$, and $\hat{e}_a^{\ B} = \delta_a^B - i\partial_a \bar{\theta} \gamma^B C \bar{\theta} - i\partial_a \bar{\lambda} \gamma^B C \bar{\lambda}$. The couplings of the $R$ axion field to these NG fields are also derived by using the supersymmetric and $R$ covariant derivative. In section IV, the dual scalar field action in the two dimensional spacetime is constructed by introducing two vectorial Lagrangian multipliers into the string action. The resulting action is found to have the form

$$\Gamma_0 = -T \int d^2 x [\sqrt{-\det(\hat{g}_{ab} - \frac{1}{\det \hat{g}} \hat{g}_{ab'} \hat{g}_{bc'} \varepsilon^{b'c} \varepsilon^{c'd} \partial_c A_i \partial_d A_i)} + \varepsilon^{ab} \partial_b A_i J_{ia}] \qquad (2)$$

where $\hat{g}_{ab} = \hat{e}_a{}^A \hat{e}_b{}^B \eta_{AB}$ is the metric of the two dimensional spacetime, and $A_i = (A_1, A_2)$ are the dual scalar fields.

## II. Transformations of Nambu-Goldstone Fields

In the four dimension target space, we consider the underlying theory with N=1 supersymmetry and $R$ symmetry. Their algebras have the following (anti)commutation relations:

$$\{Q_\alpha, \overline{Q}_{\dot{\alpha}}\} = 2\sigma^\mu_{\alpha\dot{\alpha}} P_\mu$$
$$[R, Q_\alpha] = Q_\alpha, \quad [R, \overline{Q}_{\dot{\alpha}}] = -\overline{Q}_{\dot{\alpha}}$$
$$[M_{\mu\nu}, M_{\rho\sigma}] = i(\eta_{\mu\sigma} M_{\nu\rho} + \eta_{\nu\rho} M_{\mu\sigma} - \eta_{\mu\rho} M_{\nu\sigma} - \eta_{\nu\sigma} M_{\mu\rho})$$
$$[Q_\alpha, M_{\mu\nu}] = i\frac{1}{2}(\sigma_{\mu\nu})_\alpha{}^\beta Q_\beta$$
$$[\overline{Q}^{\dot{\alpha}}, M_{\mu\nu}] = i\frac{1}{2}(\overline{\sigma}_{\mu\nu})^{\dot{\alpha}}{}_{\dot{\beta}} \overline{Q}^{\dot{\beta}} \qquad (3)$$

The coordinates of the target N=1, D=4 superspace are defined by $\{x^\mu, \theta^\alpha, \overline{\theta}_{\dot{\alpha}}\}$. Placing the vortex string along the $x$-axis, the world sheet of the vortex string is then parameterized by $\{x^0, x^1\}$ in static gauge. The related symmetries of the string defect are the Lorentz boost along the x-axis direction, i.e. the $SO(1,1)$ Lorentz symmetry (formed by the generator $M^{ab}$, $a,b=0,1$), and the rotational $SO(2)$ invariance (formed by the generator $T = M^{23}$) in the $y-z$ plane. When the target space supersymmetry and the $R$ symmetry are both spontaneously broken by the embedded vortex string, the stability subgroup $H$ of the vortex string world sheet is given by the direct product of the symmetry groups $SO(1,1) \times SO(2)$, which is spanned by the set of unbroken generators $\{M^{ab}, T\}$. These are the automorphism generators of the unbroken generators $P_a$ associated with the translational directions in D=2 spacetime. The broken automorphism generators are $K_1^a = M^{a2}$, $K_2^a = M^{a3}$ and $R$. The broken (super) spacetime generators are $Q_\alpha, \overline{Q}_{\dot{\alpha}}$ (or $q, s$, see definitions in (4)) in the superspace related to the Grassmann coordinate directions $\theta_\alpha, \overline{\theta}_{\dot{\alpha}}$ and $P_2, P_3$ related to translational directions transverse to the brane.

In D=2 dimensions, the spinor has representation of dimension $2^{2/2} = 2$, which is the same dimension as that of the D=3 spinor representation. Taking the metric tensor as $\eta^{ab} = (+,-)$ and $\gamma^a = (\sigma^2, i\sigma^1)$, the Majorana spinors $q, s$ with two real components in two dimensions are related to the D=4 Weyl spinors $Q_\alpha, \overline{Q}_{\dot{\alpha}}$ by the following relations

$$\begin{pmatrix} q_1 \\ q_2 \end{pmatrix} = \begin{pmatrix} \frac{1}{2}Q_1 e^{i\frac{\pi}{4}} - \frac{1}{2}Q_2 e^{i\frac{\pi}{4}} + \frac{1}{2}\overline{Q}_1 e^{-i\frac{\pi}{4}} - \frac{1}{2}\overline{Q}_2 e^{-i\frac{\pi}{4}} \\ \frac{1}{2}Q_1 e^{-i\frac{\pi}{4}} + \frac{1}{2}Q_2 e^{-i\frac{\pi}{4}} + \frac{1}{2}\overline{Q}_1 e^{i\frac{\pi}{4}} + \frac{1}{2}\overline{Q}_2 e^{i\frac{\pi}{4}} \end{pmatrix},$$

$$\begin{pmatrix} s_1 \\ s_2 \end{pmatrix} = \begin{pmatrix} \frac{1}{2}Q_1 e^{-i\frac{\pi}{4}} - \frac{1}{2}Q_2 e^{-i\frac{\pi}{4}} + \frac{1}{2}\overline{Q}_1 e^{i\frac{\pi}{4}} - \frac{1}{2}\overline{Q}_2 e^{i\frac{\pi}{4}} \\ \frac{1}{2i}Q_1 e^{-i\frac{\pi}{4}} + \frac{1}{2i}Q_2 e^{-i\frac{\pi}{4}} - \frac{1}{2i}\overline{Q}_1 e^{i\frac{\pi}{4}} - \frac{1}{2i}\overline{Q}_2 e^{i\frac{\pi}{4}} \end{pmatrix}. \quad (4)$$

Through dimension reduction the target spacetime symmetry can be rewritten as the centrally extended N=(2,2) supersymmetry in two dimensions, in which the number of fermionic charges is the same as that of N=1, D=4 SUSY. Thus one finds

$$\{q_i, q_j\} = 2(\gamma^a C)_{ij} P_a - 2\sigma_1 Z_1, \quad \{s_i, s_j\} = 2(\gamma^a C)_{ij} P_a - 2\sigma_1 Z_1$$

$$\{q_i, s_j\} = -2i\sigma^2 Z_2, \quad [R, q_i] = is_i, \quad [R, s_i] = -iq_i$$

$$[M^{ab}, q_i] = -\frac{1}{2}\gamma_{ij}^{ab} q_j, \quad [M^{ab}, s_i] = -\frac{1}{2}\gamma_{ij}^{ab} s_j.$$

$$[K_2^a, q_i] = \frac{1}{2}\gamma_{ij}^a s_j, \quad [K_2^a, s_i] = -\frac{1}{2}\gamma_{ij}^a q_j$$

$$[K_1^a, q_i] = -\frac{1}{2}(\sigma^3 \gamma^a)_{ij} q_j, \quad [K_1^a, s_i] = -\frac{1}{2}(\sigma^3 \gamma^a)_{ij} s_j \quad (5)$$

where $C = \gamma^0 = \sigma^2, i = 1,2$; $\gamma^{ab} = \frac{i}{2}[\gamma^a, \gamma^b], a,b = 0,1$. The central charges $Z_1$, $Z_2$ are identified with the two spatial translation generators $P_2$ and $P_3$, i.e. $Z_i = (Z_1, Z_2) = (P_2, P_3)$. The commutation relations of the generators $M$ and $T$ ($T = M^{23}$) with these of the Coset manifold are given by

$$[M, P^a] = -i\varepsilon^{ab} P_b, [M, Z_i] = 0, [M, K_i^a] = -i\varepsilon^{ab} K_{ib}, [M, R] = 0$$

$$[T, P^a] = 0, [T, Z_i] = i\varepsilon_{ij} Z_j, [T, K_i^a] = i\varepsilon_{ij} K_j^a, [T, R] = 0$$

$$[2T, \vartheta_i] = i\varepsilon_{ij}\vartheta_j, [M, T] = 0 \quad (6)$$

where $M_{ab} = -\varepsilon_{ab} M$, $\vartheta_i = (\vartheta_1, \vartheta_2) = (\sigma^3 q, s)$, $\varepsilon_{12} = 1$. The remaining communication relations are

$$[P^a, K_i^b] = -i\eta^{ab} Z_i, \quad [Z_i, K_j^a] = -iP^a \delta_{ij}, \quad [K_i^a, K_j^b] = -i\delta_{ij} M\varepsilon^{ab} - i\eta^{ab} T\varepsilon_{ij} \quad (7)$$

The Coset representative elements $\Omega = G/H = G/SO(1,1) \times SO(2)$ with respect to the stability group $H$ can be exponentially parameterized as

$$\Omega = e^{ix^a P_a} e^{i(\phi_i Z_i + \overline{\chi}_i \vartheta_i)} e^{iu_i^a K_{ia}} e^{irR}$$

$$= e^{ix^a P_a} e^{i(\phi_i(x)Z_i + \overline{\theta}_i(x)q_i + \overline{\lambda}_i(x)s_i)} e^{iu_i^a(x)K_{ia}} e^{ir(x)R} \quad (8)$$

in static gauge. The induced transformations for the collective coordinates $x^a, \phi_i, \theta_i, \lambda_i, u_i, r$ parameterizing the Coset space can be derived from the left action of full group elements on $\Omega$. Since the full symmetry group $G$ is broken down to the subgroup $H'$ (formed by the set $\{P^a, M, T\}$), a Nambu-Goldstone bosonic field corresponding to $R$ symmetry breaking is therefore expected, and it becpmes necessary to include its dynamics in the string's full actions.

Consider the transformations of the Nambu-Goldstone (Goldstino) fields induced by the left action of full group element $g$

$$g = e^{i(a^a P_a + z_i Z_i + bM + c_i^a K_{ia} + \rho T + \bar{\xi}_i q_i + \bar{\eta}_i s_i + fR)} \tag{9}$$

on the Coset representative elements $\Omega$. The result can be uniquely decomposed as the product of the new Coset $\Omega'$ and the element of $H$, i.e.

$$g\Omega = \Omega'h \tag{10}$$

where $\Omega'$ is given by

$$\Omega' = e^{ix'^a P_a} e^{i(\phi_i'(x')Z_i + \bar{\theta}_i'(x')q_i + \bar{\lambda}_i'(x')s_i)} e^{iu_i'^a(x')K_{ia}} e^{ir'(x')R} \tag{11}$$

and the stability group element can be written as

$$h = e^{i(b'M + \rho'T)} \tag{12}$$

Note that the Lie algebras of Lorentz group have the same complex extension as that of the group $SU(2) \times SU(2)$. Accordingly, after redefining generators

$$S_1 = \frac{1}{2}(T - iM), \ S_2 = \frac{1}{2}(-K_2^1 - iM), \ S_3 = \frac{1}{2}(K_1^1 + iK_2^0) \tag{13a}$$

and

$$T_1 = \frac{1}{2}(T + iM), \ T_2 = \frac{1}{2}(-K_2^1 - iK_1^0), \ T_3 = \frac{1}{2}(K_1^1 - iK_2^0) \tag{13b}$$

the following commutation relations are obtained:

$$[S_1, S_{A'}] = i\varepsilon_{A'B'}S_{B'}, \ [S_{A'}, S_{B'}] = i\varepsilon_{A'B'}S_1$$
$$[T_1, T_{A'}] = i\varepsilon_{A'B'}T_{B'}, \ [T_{A'}, T_{B'}] = i\varepsilon_{A'B'}T_1, \ [S_{1(A')}, T_{1(B')}] = 0 \tag{14}$$

where $A', B' = 2, 3$, $\varepsilon_{A'B'}\varepsilon_{A'C'} = \delta_{B'C'}$, and $\varepsilon_{23} = -\varepsilon_{32} = 1$. Therefore, Eq.(10) becomes

$$e^{i(a^a P_a + z_i Z_i + b_1 S_1 + \rho_1 T_1 + m_A S_{A'} + n_A T_{A'} + \bar{\xi}_i q_i + \bar{\eta}_i s_i + fR)} e^{ix^a P_a} e^{i(\phi_i Z_i + \bar{\theta}_i q_i + \bar{\lambda}_i s_i)} e^{i(u_A S_{A'} + v_B T_{B'})} e^{irR}$$
$$= e^{ix'^a P_a} e^{i(\phi_i' Z_i + \bar{\theta}_i' q_i + \bar{\lambda}_i' s_i)} e^{i(u_A' S_{A'} + v_B' T_{B'})} e^{ir'R} e^{i(b_1' S_1 + \rho_1' T_1)} \tag{15}$$

Applying Baker-Campbell-Hausdorff formulas extensively to Eq.(15), up to the first order of the parameters of $g$, leads to the induced general infinitesimal coordinate transformations along with those of the Nambu-Goldstone (Goldstino) fields on the two dimensional world sheet:

$$x'^a = x^a + a^a + i\bar{\xi}_i(\gamma^a C)_{ij}\bar{\theta}_j + i\bar{\eta}_i(\gamma^a C)_{ij}\bar{\lambda}_j + \varepsilon^{a'a}x_a b - c_i^a \phi_i$$
$$\quad + i2\bar{\theta}_i(\gamma^a C)_{ij}\bar{\xi}_j + i2\bar{\lambda}_i(\gamma^a C)_{ij}\bar{\eta}_j$$

$$\phi_1' = \phi_1 + z_1 - i\bar{\xi}_i(\sigma_1)_{ij}\bar{\theta}_j - i\bar{\eta}_i(\sigma_1)_{ij}\bar{\lambda}_j - c_1^a x_a + \rho\phi_2 - i2\bar{\theta}_i(\sigma_1)_{ij}\bar{\xi}_j - i2\bar{\lambda}_i(\sigma_1)_{ij}\bar{\eta}_j$$

$$\phi_2' = \phi_2 + z_2 - \bar{\theta}_i(\sigma^2)_{ij}\bar{\eta}_j + \bar{\xi}_i(\sigma^2)_{ij}\bar{\lambda}_j - c_2^a x_a - \rho\phi_1 - 2\bar{\xi}_i(\sigma^2)_{ij}\bar{\lambda}_j + 2\bar{\theta}_i(\sigma^2)_{ij}\bar{\eta}_j$$

$$\bar{\theta}_i' = \bar{\theta}_i + \bar{\xi}_i + f\bar{\lambda}_i - i\frac{1}{2}b\bar{\theta}_j\varepsilon_{ab}\gamma_{ji}^{ab} - i\frac{1}{2}c_1^a\bar{\theta}_j(\sigma^3\gamma^{a'})_{ji}\eta_{aa'} - i\frac{1}{2}c_2^a\bar{\lambda}_j\gamma_{ji}^{a'}\eta_{aa'} - i\frac{1}{2}\rho\bar{\lambda}_j\gamma_{ji}^2$$

$$\bar{\lambda}_i' = \bar{\lambda}_i + \bar{\eta}_i - f\bar{\theta}_i - i\frac{1}{2}b\bar{\lambda}_j\varepsilon_{ab}\gamma_{ji}^{ab} + i\frac{1}{2}c_2^a\bar{\theta}_j\gamma_{ji}^{a'}\eta_{aa'} - i\frac{1}{2}c_1^a\bar{\lambda}_j(\sigma^3\gamma^{a'})_{ji}\eta_{aa'} + i\frac{1}{2}\rho\bar{\theta}_j\gamma_{ji}^2$$

$$r' = r + f \tag{16}$$

As a result, the Nambu-Goldstino fields are found to transform inhomogeneously, and these inhomogeneous terms explicitly signal the breakdown of the supersymmetry. As for the spacetime symmetry breaking, the Nambu-Goldstone fields are those associated with the broken (super)translations only [23]. The Nambu-Goldstone fields $u_i^a$ are actually superfluous (non-dynamic) and can be eliminated by the inverse Higgs mechanism [24] (also see Eq.(32)). Their transformations, however, can be obtained by considering the infinitesimal transformations of the Nambu-Goldstone fields $u_{A'}, v_{B'}$ and $b_1, \rho_1$ through Eq.(15), i.e.

$$u_{A'}' = u_{A'} + m_{A'} - \frac{1}{2}b_1 u_{B'}\varepsilon_{B'A'} + \frac{\coth(i\frac{1}{2}\sqrt{u \cdot u})}{u \cdot u}(i\frac{1}{2}\sqrt{u \cdot u})u \cdot m u_{B'}\varepsilon_{B'A'}$$

$$- \frac{u \cdot m u_{B'}\varepsilon_{B'A'}}{u \cdot u} + \frac{1}{2}b_1' u_{B'}\varepsilon_{B'A'}$$

$$v_{B'}' = v_{B'} + n_{B'} - \frac{1}{2}\rho_1 v_{A'}\varepsilon_{A'B'} + \frac{\coth(i\frac{1}{2}\sqrt{v \cdot v})}{v \cdot v}(i\frac{1}{2}\sqrt{v \cdot v})v \cdot n v_{A'}\varepsilon_{A'B'}$$

$$- \frac{v \cdot n v_{A'}\varepsilon_{A'B'}}{v \cdot v} + \frac{1}{2}\rho_1' v_{A'}\varepsilon_{A'B'} \tag{17a}$$

and

$$b_1' = b_1 + \frac{\frac{1}{2}u \cdot m}{\coth(i\frac{1}{2}\sqrt{u_{A'}^2})(i\frac{1}{2}\sqrt{u_{A'}^2})} ; \quad \rho_1' = \rho_1 + \frac{\frac{1}{2}v \cdot n}{\coth(i\frac{1}{2}\sqrt{v_{A'}^2})(i\frac{1}{2}\sqrt{v_{A'}^2})} \tag{17b}$$

where $u \cdot m = u_{A'} m_{B'}\varepsilon_{A'B'}$, $u_{A'}^2 = u_{A'} u_{A'}$ etc., and $u_{A'}, v_{B'}, m_{A'}, n_{B'}, b_1, \rho_1$ are related to $u_i^a, c_i^a, b, \rho$ through Eqs.(10,13,15) up to field redefinitions.

## III. Effective Actions

As a result of Eqs.(16,17), the NG fields nonlinearly spontaneous break the supersymmetric and $R$ symmetries. The effective actions of the string can be constructed by using the Zweibein and connection one-forms from the Coset structure $\Omega^{-1}d\Omega$, which can be explicitly expanded with respect to the $G$ generators as

$$\Omega^{-1}d\Omega = i(\omega^A P_A + \overline{\omega}_{\vartheta_i}\vartheta_i + \omega_{z_i}Z_i + \omega_{k_i}^A K_{iA} + \omega_T T + \omega_M M + \omega_R R)$$
$$= i(\omega^A P_A + \overline{\omega}_{q_i}q_i + \overline{\omega}_{s_i}s + \omega_{z_i}Z_i + \omega_{k_i}^A K_{iA} + \omega_T T + \omega_M M + \omega_R R) \tag{18}$$

Under the transformation of Eq.(10) the Coset transforms as $\Omega \to \Omega'$, and the Maurer-Cartan one-forms transform according to

$$\Omega'^{-1}d\Omega' = h(\Omega^{-1}d\Omega)h^{-1} + hdh^{-1}. \tag{19}$$

Obviously, all the one-forms transform homogeneously under $G$ except the connection one-forms $\omega_M$ and $\omega_T$, which transform by a shift. The covariant coordinate one-forms $\omega^A$ transform as the Zweibein $e_b{}^A$ on the tangent bundle to $G/H$, while $\omega_M, \omega_T$ transform as connections to this bundle. The Zweibein $e_b{}^A$ can be obtained by expanding the covariant coordinate one-forms $\omega^A$ with respect to the general coordinate differentials $dx^b$, i.e. $\omega^A = dx^b e_b{}^A$. On the other hand, the connection one-forms $\omega_M$ and $\omega_T$ can be used to construct the covariant derivatives for the fields localized on the brane world volume [25]. These are the building blocks that can be used to construct the invariant actions. Applying the differential formulas $\exp(-b)d\exp(b) = \sum_{k=0}^{\infty} \frac{(-1)^k}{(k+1)!}(ad_b)^k db$ to Eq.(18), where $ad_{b/2}(a) = [\frac{b}{2}, a]$, gives us

$$i\omega^A p_A = (idx^B + d\bar{\theta}_i(\gamma^B C)_{ij}\theta_j + d\bar{\lambda}_i(\gamma^B C)_{ij}\lambda_j)P_A(\delta_B{}^A + u_{iB}(U^{-1}(\cosh\sqrt{U}-1))_{ij}u_j^A)$$

$$+ i(d\phi_i + d\psi_i)(U^{-\frac{1}{2}}\sinh\sqrt{U})_{ij}u_j^A P_A$$

$$i\omega_{z_i} Z_i = i(d\phi_j + d\psi_j)(\cosh\sqrt{U})_{ji} Z_i$$

$$+ (idx^B + d\bar{\theta}\gamma^B C\bar{\theta} + d\bar{\lambda}\gamma^B C\bar{\lambda})u_{jB}(\cosh\sqrt{U}U^{-\frac{1}{2}}\tanh\sqrt{U})_{ji} Z_i \quad (20)$$

where $x^B = (x^0, x^1)$, the matrix $U_{ij} = u_i^a u_{ja}$, and

$$d\psi_1 = i(d\bar{\theta}\sigma_1\bar{\theta} + d\bar{\lambda}\sigma_1\bar{\lambda})$$
$$d\psi_2 = \bar{\theta}\sigma^2 d\bar{\lambda} - d\bar{\theta}\sigma^2\bar{\lambda} \quad (21)$$

in which $A, B = 0, 1$. The capital letters $A, B$ are used to represent the covariant spacetime coordinate indices, and the lowercase letters $a, b$ are used to represent $1+1$ general coordinate indices in what follows. Considering $\omega^A = d\xi^b e_b{}^A = dx^b e_b{}^A$ in static gauge, the Zweibein is therefore found to be

$$e_a{}^A = (\delta_a{}^B - i\partial_a\bar{\theta}\gamma^B C\theta - i\partial_a\bar{\lambda}\gamma^B C\lambda)(\delta_B{}^A + u_{iB}(U^{-1}(\cosh\sqrt{U}-1))_{ij}u_j^A$$

$$+ (\partial_a\phi_i + \partial_a\psi_i)(U^{-\frac{1}{2}}\sinh\sqrt{U})_{ij}u_j^A)$$

$$= \hat{e}_a{}^B(\delta_B{}^A + u_{iB}(U^{-1}(\cosh\sqrt{U}-1))_{ij}u_j^A + \nabla_B\phi_i(U^{-\frac{1}{2}}\sinh\sqrt{U})_{ij}u_j^A) \quad (22)$$

in which $\nabla_B\phi_i = \hat{e}_B^{-1a}(\partial_a\phi_i + \partial_a\psi_i)$, and $\hat{e}_B^{-1a}\partial_a$ is the Akulov-Volkov derivative, defined by $\hat{e}_a{}^B = \delta_a^B - i\partial_a\bar{\theta}\gamma^B C\bar{\theta} - i\partial_a\bar{\lambda}\gamma^B C\bar{\lambda}$ [26]. Under the transformations of Eq.(19), the covariant coordinate differentials transform as

$$i\omega'^A P'_A = idx'^a e'_a{}^A P'_A$$
$$= e^{i(bM+\rho T)}idx^b e_b{}^B P_B e^{-i(bM+\rho T)} = idx^b e_b{}^B L_B{}^A P_A \quad (23)$$

where $L_B{}^A$ is the local $H$ representation with vector indices. As a result, the transformation of Zweibein induced by Eq.(23) has the property

$$e'_a{}^A = \frac{\partial x^b}{\partial x'^a} e_b{}^B L_B{}^A \quad (24)$$

In addition, $P_A$ is invariant under the action of $SO(2)$, therefore the induced total variation of Zweibein is a $SO(2)$ scalar, i.e.

$$\delta_{SO(2)}(e_a{}^B) = 0 \qquad (25)$$

Obviously, under the $R$ transformation (16), the total variation of Zweibein is $R$ invariant according to Eq.(19):

$$\delta_R(e_a{}^B) = 0 \qquad (26)$$

In order to construct the action of the $R$ axion related to the broken $R$ symmetry, from Eq.(18) one finds

$$i\omega_R R = idrR = idx^a \partial_a rR = i\omega^A e_A^{-1a} \partial_a rR \qquad (27)$$

The covariant derivative is then given by

$$\nabla_A r = e_A^{-1a} \partial_a r \qquad (28)$$

in which it has a covariant coordinate index $A$. Under the transformation of Eq.(16) it is also $R$ invariant

$$\delta_R(\nabla_A r) = 0 \qquad (29)$$

Furthermore, in Eq.(22) one may note that there are no derivative terms for the Nambu-Goldstone fields $u_i^a$. As a result, they have no dynamical degrees of freedom. These superfluous fields can be eliminated by using their equations of motion or by imposing covariant constraints on the system. Imposing $\omega_z = 0$ as the covariant constraint on Eq.(20) yields

$$\begin{aligned}\omega_{z_i} &= (d\phi_j + d\psi_j)\,(\cosh\sqrt{U})_{ji} \\ &\quad + (dx^B - id\bar{\theta}\gamma^B C\theta - id\bar{\lambda}\gamma^B C\lambda)u_{jB}\,(\cosh\sqrt{U}\,U^{-\frac{1}{2}}\tanh\sqrt{U})_{ji} \\ &= dx^a(\delta_a{}^B - i\partial_a\bar{\theta}\gamma^B C\theta - i\partial_a\bar{\lambda}\gamma^B C\lambda)(\nabla_B\phi_j\,(\cosh\sqrt{U})_{ji} \\ &\quad + u_{jB}\,(\cosh\sqrt{U}\,U^{-\frac{1}{2}}\tanh\sqrt{U})_{ji}) \\ &= dx^a \hat{e}_a{}^B\,(\nabla_B\phi_j\,(\cosh\sqrt{U})_{ji} + u_{jB}(\cosh\sqrt{U}\,U^{-\frac{1}{2}}\tanh\sqrt{U})_{ji}) \\ &= 0 \end{aligned} \qquad (30)$$

Hence one can find

$$\nabla_B \phi_i = -u_{jB}\,(U^{-\frac{1}{2}}\tanh\sqrt{U})_{ji}, \qquad (31)$$

or inversely,

$$u_{iB} = -\nabla_B \phi_j\,(U^{-\frac{1}{2}}\tanh\sqrt{U}))^{-1}{}_{ji}. \qquad (32)$$

Introduce a new quantity

$$H_{kk'} = \nabla_B \phi_k \nabla^B \phi_{k'} = (\tanh\sqrt{U})^2{}_{kk'} \qquad (33)$$

Plugging Eq.(32) into $U_{ij} = u_i^a u_{ja}$ amounts to

$$U_{ij} = (U^{-\frac{1}{2}}\tanh\sqrt{U})^{-1}{}_{ik}\nabla_B\phi_k \nabla^B\phi_{k'}(U^{-\frac{1}{2}}\tanh\sqrt{U})^{-1}{}_{k'j}$$

$$= (U^{-\frac{1}{2}} \tanh\sqrt{U})^{-1}{}_{ik} H_{kk'} (U^{-\frac{1}{2}} \tanh\sqrt{U})^{-1}{}_{k'j} \tag{34}$$

The Zweibein is secured as

$$e_a{}^A = \hat{e}_a{}^B (\delta_B{}^A + u_{iB}(U^{-1}(\cosh\sqrt{U}-1))_{ij} u_j^A + \nabla_B\phi_i (U^{-\frac{1}{2}}\sinh\sqrt{U})_{ij} u_j^A)$$

$$= \hat{e}_a{}^B (\delta_B{}^A + \nabla_B\phi_i H_{ij}^{-\frac{1}{2}} (\cosh\sqrt{U}-1-\tanh\sqrt{U}\sinh\sqrt{U})_{jk} H_{kl}^{\frac{1}{2}} \nabla^A\phi_l)$$

$$= \hat{e}_a{}^B (\delta_B{}^A + \nabla_B\phi_i H_{ij}^{-\frac{1}{2}} (\sqrt{1-H}-1)_{jk} H_{kl}^{\frac{1}{2}} \nabla^A\phi_l) \tag{35}$$

By using Eq.(35) the metric tensor of the two dimensional world sheet becomes

$$g_{ab} = e_a{}^A e_b{}^B \eta_{AB}$$

$$= \hat{e}_a{}^{A'} \hat{e}_b{}^{B'} (\eta_{A'B'} - \nabla_{A'}\phi_i \nabla_{B'}\phi_i) \tag{36}$$

The transformation properties of $i\phi_i Z_i$ under the action of $H$ can be concluded by considering Eqs.(6,19). It follows that $\phi_i$ is a $SO(2)$ vector but a $SO(1,1)$ scalar. Therefore the total variation is

$$\delta_{SO(2)}(\nabla_{A'}\phi_i \nabla_{B'}\phi_i) = 0. \tag{37}$$

Considering Eq.(16), one can show it is also $R$ invariant

$$\delta_R(\nabla_{A'}\phi_i \nabla_{B'}\phi_i) = 0 \tag{38}$$

Besides, in the case of $R$ variation of $\hat{e}_a{}^B = \delta_a{}^B - i\partial_a\bar{\theta}\gamma^B C\bar{\theta} - i\partial_a\bar{\lambda}\gamma^B C\bar{\lambda}$, after plugging in Eq.(16) one finds

$$\delta_R(\hat{e}_a{}^B) = 0 \tag{39}$$

Under the general coordinate transformations of Eq.(16), by using Eq.(24) one can find the following invariance

$$d^2x'\sqrt{|\det g'|} = d^2x' \det e'_b{}^A = d^2x \det\left|\frac{\partial x'^b}{\partial x^a}\right| \det\left|\frac{\partial x^b}{\partial x'^a}\right| \det e_b{}^B \det L_B{}^A$$

$$= d^2x \det e_b{}^B = d^2x\sqrt{|\det g|} \tag{40}$$

Therefore, considering Eqs.(37,38,39), the $SO(1,1)\times SO(2)$ and $R$ invariant effective action of the vortex string is found to be

$$\Gamma_0 = -T\int d^2x \sqrt{|\det g|} = -T\int d^2x \det\hat{e}\sqrt{\det(\eta_{A'B'} - \nabla_{A'}\phi_i \nabla_{B'}\phi_i)}$$

$$= -T\int d^2x \det\hat{e}\sqrt{\det(\delta_{ij} - \nabla_A\phi_i \nabla^A\phi_j)} \tag{41}$$

As a result, it follows that the string action takes a factorized form of Akulov-Volkov and Nambu-Goto actions. Moreover, considering Eqs.(28,29), the action of the $R$ axion field, which is invariant under both the supersymmetric and $R$ symmetries, is given by

$$\Gamma_R = -T_R\int d^2x \sqrt{|\det g|}\ell_R = -T_R\int d^2x \det e \nabla^A r \nabla_A r \tag{42}$$

Consequently, the total effective actions of the vortex string describing its long wave oscillations into the target (super)space as well as the internal $R$ space are found to be

$$\Gamma = \Gamma_0 + \Gamma_R$$

$$= -T\int d^2x \det\hat{e}\sqrt{\det(\delta_{ij} - \nabla_A\phi_i \nabla^A\phi_j)} - T_R\int d^2x \det e \nabla^A r \nabla_A r \tag{43}$$

where $\det\hat{e}$ is the determent of $\hat{e}_a{}^B$, and $T$ stands for string tension. $T_R$ is related to the coupling of the $R$ axion field to other Nambu-Goldstone(Goldstino) fields.

If instead the $R$ symmetry is not broken, then the stability subgroup takes the form $H_1 = SO(1,1) \times SO(2) \times R$ which is spanned by the set of unbroken generators $\{M,T,R\}$. Such a p=1 brane topological defect then breaks the full $G$ symmetry down to the subgroup $H_1{}'$ (formed by the set $\{M,P_a,T,R\}$). As a result, its Coset becomes $\Omega_1 = G/H_1 = G/SO(1,1) \times SO(2) \times R$, whileas the representative elements are parameterized as

$$\Omega_1 = e^{ix^a P_a} e^{i(\phi_i(x) Z_i + \bar{\chi}_i(x) \vartheta_i)} e^{iu_i^a(x) K_{ia}}$$
$$= e^{ix^a P_a} e^{i(\phi_i(x) Z_i + \bar{\theta}_i(x) q_i + \bar{\lambda}_i(x) s_i)} e^{iu_i^a(x) K_{ia}} \quad (44)$$

in static gauge.

By means of Eq.(8), one finds

$$\Omega = \Omega_1 e^{irR}. \quad (45)$$

It further leads to

$$\Omega^{-1} d\Omega = e^{-irR} \Omega_1^{-1} d(\Omega_1 e^{irR})$$
$$= e^{-irR} (\Omega_1^{-1} d\Omega_1) e^{irR} + e^{-irR} de^{irR}$$
$$= e^{-irR} i(\omega_1^A P_A + \bar{\omega}_{1q_i} q_i + \bar{\omega}_{1s_i} s_i + \omega_{1z_i} Z_i + \omega_{1k_i}^A K_{iA} +$$
$$\omega_{1T} T + \omega_{1M} M + \omega_{1R} R) e^{irR} + e^{-irR} de^{irR}$$
$$= i(\omega^A P_A + \bar{\omega}_{q_i} q_i + \bar{\omega}_{s_i} s_i + \omega_{z_i} Z_i + \omega_{k_i}^A K_{iA} + \omega_T T + \omega_M M + \omega_R R) \quad (46)$$

It can be shown that

$$\omega^A = \omega_1^A; \omega_{z_i} = \omega_{1z_i};$$
$$\bar{\omega}_{q_i} \cosh(-ir) - i\bar{\omega}_{s_i} \sinh(-ir) = \bar{\omega}_{1q_i};$$
$$i\bar{\omega}_{q_i} \sinh(-ir) + \bar{\omega}_{s_i} \cosh(-ir) = \bar{\omega}_{1s_i};$$
$$\omega_{k_i}^A = \omega_{k_i}^A; \omega_T = \omega_{1T}; \omega_M = \omega_{1M};$$
$$\omega_R = \omega_{1R} + idrR = idrR. \quad (47)$$

Since $R$ is the automorphism generator of the supercharges and commutes with the Poincare group generators, it is obvious that the $R$ symmetry one-forms vanishes in $\Omega_1^{-1} d\Omega_1$, i.e. $i\omega_{1R} R = 0$. Furthermore, it can be shown that the covariant coordinate one-forms $\omega_1^A = \omega^A$ is also $R$ invariant when $\Omega_1 \to \Omega_1{}'$. In this context, in accordance with the expansion $\omega_1^A = dx^a e_{1a}{}^A$, the effective action of the string is purely described by Eq.(41), i.e., the invariant synthesis form of the Akulov-Volkov and Nambu-Goto actions:

$$\Gamma_0 = -T \int d^2 x \det e_1$$
$$= -T \int d^2 x \det \hat{e} \sqrt{\det(\delta_{ij} - \nabla_A \phi_i \nabla^A \phi_j)} \quad (48)$$

It is possible for the matter and gauge fields to be trapped on the brane in the presence of topological defects [27, 28], and the localized matter degrees of freedom on the brane world volume can be introduced by using the covariant derivative of Eq.(28). In the action of Eq.(43), the coupling constant $T_R$ is related to the $R$ symmetry breaking scale, which might be different from the scale of SUSY breaking. As a result, the action of the string, which is actually described by the Nambu-Goldstone modes corresponding to its low energy oscillations into the covolume (super)space, can be reinterpreted as an effective two dimensional field theory, in which the otherwise possible SUSY breaking terms have been integrated out.

## IV. Dual Scalar Field Theory

From the two dimensional point of view the effective action of the vortex string has a factorized form of Akulov-Volkov-Nambu-Goto actions. In what follows, a form of dual scalar field theory is introduced. The dual action is found to describe the dynamics of some scalar fields localized on the world volume, and whose metric is given by $\hat{g}_{ab} = \hat{e}_a{}^A \hat{e}_b{}^B \eta_{AB}$. To begin we start with the string action

$$\Gamma_0 = -T \int d^2 x \det \hat{e} \sqrt{\det(\delta_{ij} - \nabla_A \phi_i \nabla^A \phi_j)}$$
$$= -T \int d^2 x \det \hat{e} \sqrt{1 - \nabla^A \phi_1 \nabla_A \phi_1 - \nabla^A \phi_2 \nabla_A \phi_2 + (\nabla \phi_1)^2 (\nabla \phi_2)^2 - (\nabla^A \phi_1 \nabla_A \phi_2)^2} \quad (49)$$

Introducing two vectorial Lagrangian multipliers $L_i^A$ into the action gives us

$$\Gamma_0 = -T \int d^2 x [\det \hat{e} \sqrt{1 - l_i^2 + l_1^2 l_2^2 - (l_1 l_2)^2} + \det \hat{e} L_i^A (l_{iA} - \nabla_A \phi_i)] \quad (50)$$

where $l_{iA} = \nabla_A \phi_i$, $l_i^2 = \nabla_A \phi_i \nabla^A \phi_i$. Applying the equations of motion of $\phi_i$ amounts to

$$\partial_a (\det \hat{e} \hat{e}_A^{-1a} L_i^A) = 0. \quad (51)$$

Introduce two vector density quantities $F_i^a$ with weight $-1$

$$F_i^a = \det \hat{e} \hat{e}_A^{-1a} L_i^A \quad (52)$$

Considering Eq.(51), their solutions are found to be

$$F_i^a = \varepsilon^{ab} \partial_b A_i \quad (53)$$

where the fields $A_i$ transform as scalars in the 1+1 dimensional spacetime. Define the covariant vectors

$$F_{ia} = \frac{1}{\det \hat{e}} \hat{g}_{ab} F_i^b = \hat{g}_{ab} \hat{e}_A^{-1b} L_i^A \quad (54)$$

Therefore one has

$$L_{iB} L_j^B = F_{ia} F_{jb} \hat{g}^{ab} \quad (55)$$

Note that

$$\det \hat{e} L_i^A \nabla_A \phi_i = \det \hat{e} L_i^A \hat{e}_A^{-1a} (\partial_a \phi_i + \partial_a \psi_i) \quad (56)$$

Considering Eq.(51) and integrating by parts, the $\phi_i$ dependent term in Eq.(56) can be eliminated form the action. Hence, Eq.(50) becomes

$$\Gamma_0 = -T \int d^2 x [\det \hat{e} \sqrt{1 - l_i^2 + l_1^2 l_2^2 - (l_1 l_2)^2} + \det \hat{e} L_i^A (l_{iA} - \hat{e}_A^{-1a} \partial_a \psi_i)] \quad (57)$$

in which $\psi_i$ are given by Eq.(21). Solving for the equations of motion of $l_{iA}$ leads to

$$L_1^A = \frac{l_1^A - l_1^A l_2^2 + l_1 l_2 l_2^A}{\sqrt{1 - l_i^2 + l_1^2 l_2^2 - (l_1 l_2)^2}}, \quad L_2^A = \frac{l_2^A - l_2^A l_1^2 + l_1 l_2 l_1^A}{\sqrt{1 - l_i^2 + l_1^2 l_2^2 - (l_1 l_2)^2}} \tag{58}$$

Then the complete dual form of action (57) has the following form

$$\Gamma_0 = -T \int d^2 x \det \hat{e} [\sqrt{(1 + L_1^2)(1 + L_2^2) - (L_1 \cdot L_2)^2}$$
$$+ L_1^A \hat{e}_A^{-1a} i(-\partial_a \bar{\theta} \sigma_1 \theta - \partial_a \bar{\lambda} \sigma_1 \lambda) + L_2^A \hat{e}_A^{-1a} (-\bar{\theta} \sigma^2 \partial_a \bar{\lambda} + \partial_a \bar{\theta} \sigma^2 \bar{\lambda})] \tag{59}$$

By using Eq.(52) it can be shown that $L_i^A = F_i^a \det^{-1} \hat{e} \hat{e}_a^A$. Define

$$J_{1a} = -i(\partial_a \bar{\theta} \sigma_1 \theta + \partial_a \bar{\lambda} \sigma_1 \lambda), \quad J_{2a} = -(\bar{\theta} \sigma^2 \partial_a \bar{\lambda} - \partial_a \bar{\theta} \sigma^2 \bar{\lambda}) \tag{60}$$

The action (59) is converted to

$$\Gamma_0 = -T \int d^2 x [\det \hat{e} \sqrt{(1 + L_1^2)(1 + L_2^2) - (L_1 \cdot L_2)^2} + F_i^a J_{ia}]$$
$$= -T \int d^2 x [\det \hat{e} \sqrt{\det(\delta_{ij} + L_i L_j)} + F_i^a J_{ia}]$$
$$= -T \int d^2 x [\det \hat{e} \sqrt{\det(\delta_{ij} + F_{ia} F_{jb} \hat{g}^{ab})} + F_i^a J_{ia}]$$
$$= -T \int d^2 x [\sqrt{-\det \hat{g}} \sqrt{\det(\delta_a^b + F_{ia} F_{ic} \hat{g}^{cb})} + F_i^a J_{ia}] \tag{61}$$

where $\hat{g}_{ab} = \hat{e}_a^{\ A} \hat{e}_b^{\ B} \eta_{AB}$. Therefore the action that describes the dynamics of the scalar fields $A_i$ localized on the world sheet with the metric $\hat{g}_{ab} = \hat{e}_a^{\ A} \hat{e}_b^{\ B} \eta_{AB}$ is secured as

$$\Gamma_0 = -T \int d^2 x [\sqrt{-\det(\hat{g}_{ab} + F_{ia} F_{ib})} + F_i^a J_{ia}]$$
$$= -T \int d^2 x [\sqrt{-\det(\hat{g}_{ab} - \frac{1}{\det \hat{g}} \hat{g}_{ab'} \hat{g}_{bc'} \varepsilon^{b'c} \varepsilon^{c'd} \partial_c A_i \partial_d A_i)} + \varepsilon^{ab} \partial_b A_i J_{ia}] \tag{62}$$

As a result, this action, as a two dimensional field theory, gives us a scalar field theory that is dual to the action of Eq.(41).

On the other hand, if one turns to the geometrical structure of the vortex string, the sting's geometrical and physical characteristics would be inevitably involved in the action. For example, the action could be supplemented with an extrinsic curvature coupling term as embedded in a higher dimensional target spacetime [29]. Furthermore, in the context of spontaneous symmetry breaking due to the non-abelian vortex string, the inclusion in the effective action of the NG modes arising from oscillations in the internal space would become natural and necessary. In such a case, in addition to the position moduli space discussed here, the moduli space of the vacuum is enlarged to include the internal Coset space of G/H, where G is the global internal symmetry of the theory, and H is the symmetry which leaves the vacuum invariant [6,7,10,11]. Therefore, the string dynamics, whose moduli space includes both spacetime and internal space symmetry breaking, would be of interest to the general theory of vortex string and is worthy of further exploration and investigation.

This work was supported in part by Purdue PRF grant. The author thanks T. ter Veldhuis, T.K.Kuo and T.E.Clark for discussions and support. The author also thanks M. Kruczenski, M.Nitta and C.Codrington for helpful comments and suggestions.